\documentstyle[manuscript, aps]{revtex}
 {\par\noindent{\underline{Proof} \quad}}{\hfill$\Box$\bigskip}
 {\par\noindent{\underline{Proof} of the theorem\quad}}{\hfill$\Box$\bigskip}
 {\par\smallskip\noindent{\underline{{\it Remark}} \quad}}{\par\smallskip}
 {\par\smallskip\noindent{\underline{{\it Fact}} \quad}}{\par\smallskip}
 {\par\smallskip\noindent{\underline{{\it Example}} \quad}}{\par\smallskip}
 {\par\smallskip\noindent{{\it Assumotion} \quad}}{\par\smallskip}
 {\par\smallskip\noindent{{\it Condition} \quad}}{\par\smallskip}

\begin{document}
\title{Detecting the inseparability and distillability of continuous variable  states in Fock space}
\author{Wang Xiang-Bin\thanks{Correspondence author, email: wang@qci.jst.go.jp}, 
Matsumoto Keiji\thanks{email: keiji@qci.jst.go.jp} and Tomita Akihisa\thanks{email: a-tomita@az.jp.nec.com}\\
       IMAI Quantum Computation and Information Project, \\ERATO,Japan Sci. and Tech. Corp.
\\Daini Hongo White Bldg. 201, 5-28-3, Hongo, BunKyo
\\ Tokyo 113-0033, Japan}

\maketitle
\begin{abstract}
The partial transposition(PT) operation is an effecient tool
in detecting the inseparability of a mixed state. 
We give an explicit formula for the PT operation for the 
continuous variable states in Fock space. 
We then give the necessary and sufficient condition
for the positivity of Gaussian operators. Based on this, a number of
creterions on the inseparability and distillability for the multimode Gaussian
states are naturally drawn. We finally give an explicit formula for the state in a subspace
of a global Gaussian state. This formula, together with the known results for Gaussian states, 
gives the criterions
for the inseparability and distillability in  a subspace of the global Gaussian  state.  
\end{abstract}
Quantum entanglement or inseparability plays a fundamentally important role in the 
fascinating area of  quantum computation and quantum information. 
To check
the inseparability of a (mixed) state, a particularly elegant mathematical
tool, partial transposition(PT) operation was given by Peres\cite{peres}. 
This method was later on  applied
to various discrete states\cite{horo1,horo2,lew}. 
While a lot of studies in quantum 
computation and quantum information(QCI) are based on the discrete states, 
recently a number of new proposals in the area
were raised by using the continuous variable states\cite{c1,c2,c3,c4,got}. 
In certain respects, the continuous variable states can have
advantages to the discrete states.  Consequently,  the entanglement detection for 
continuous variable states is an interesting and important topic. 
The distillability and  and inseparability for Gaussian states were broadly
studied\cite{lu1,simon,lu2,lu3,dist} recently.
In these works, the 
 starting point is the Wigner function, whose PT operation is simple and clear.
Actually, the PT operation in Fock space is also simple. 
 In this letter, we directly use Peres$'$
PT operation to the density operator itself. We shall first give an explicit formula for the PT
operation for the continuous variable states in Fock space. 
 We then check the inseparability of various
types of practically existing states. We formulate a new type of
practical impure squeezed state
in the form of mixture of different squeezed vacuum states and Fock number states. 
We show this state and the squeezed state in phase damping channel are always inseparable.
We then apply the PT operation to the
Gaussian operator. We shall give the general criterion
on the positivity of multi-variable Gaussian states in Fock space. 
This can be used to give the necessary and sufficient condition 
for inseparability or distillability for
$N\otimes 1$ system\cite{wern} or arbitrary bipartite system\cite{dist}, respectively.
More generally, with the positivity condition for Gaussian state in Fock space
we can easily
investigate the entanglement prpoerty in a subspace of a global Gaussian state. 
\\   
{\bf The PT operation in Fock space.} 
We start from the number state $|{n_i}\rangle $ in Fock space, $i$ from 
$1$ to $l$.   
Without loss of any generality, operators in Fock space 
can be described  by the summation of $l$-mode ket bra operator
\begin{eqnarray}
\rho=\sum_{\{n_i\}}c_{\{n_i,n'_i\}}|n_1,n_2\cdots n_l\rangle \langle  n_1',n_2'\cdots n_l'|+h.c..
\end{eqnarray}  
Suppose mode $1$ to $j-1$ belong to subspace $A$ and mode $j$ to $l$ belong to subspace $B$. 
The PT operation to subspace $B$ is 
\begin{eqnarray}
\rho^{PT}=\sum_{\{n_i\}}c_{\{n_i,n'_i\}}|n_1,n_2\cdots n_{j-1}, n_j',n_{j+1}'\cdots n_l'\rangle 
\langle  n_1',n_2'\cdots n_{j-1}', n_j, n_{j+1},\cdots, n_l| +h.c..
\label{fock1}\end{eqnarray}
An operator in Fock space can also
be expressed in creation($a^\dagger$) and annihilation($a$) operators. 
Considering the fact
$
a^\dagger=\sum^\infty _{n=0} \sqrt{n+1}|n+1\rangle \langle  n|
$ and
$
a=\sum_{n=0} ^\infty \sqrt{n+1}|n\rangle \langle  n+1|,
$
we have
 $
{a^k}^T={a^\dagger}^k
$; ${{a^\dagger}^k}^T=a^k,
$ where
$k$ is an arbitrary natural number,
the superscript $T$ indicates the $transpose$ operation. 
Without any loss of generality, an $l$-mode density operator  in Fock space can be written in the 
following normally ordered form
(i.e., all creation operators in the left and all annihilation operators in the right) 
 \begin{eqnarray}\label{state}
\hat W=f(a_1^\dagger,a_1,a_2^\dagger, a_2\cdots a_l^{\dagger},a_l)=
\sum_{\mu_i,\nu_i}
c_{\mu_1,\nu_1\cdots\mu_l,\nu_l}{a_1^{\dagger}}^{\mu_1}a_1^{\nu_1}
\cdots {a_l^{\dagger}}^{\mu_l}a_l^{\nu_l} +h.c.
\end{eqnarray}
Note that the operators in the different mode are commute to each other.
If we take the PT operation to operator $\hat W$ on the mode from $j$ to $l$, ${\hat W}$ changes
into
\begin{eqnarray}
{\hat W}^{PT}=\sum_{\mu_i,\nu_i}
c_{\mu_1,\nu_1\cdots\mu_l,\nu_l}{a_1^\dagger}^{\mu_1}a_1^{\nu_1}
\cdots {a_{j-1}^\dagger}^{\mu_{j-1}}{a_{j-1}}^{\nu_{j-1}} {a_j^\dagger}^{\nu_j}
a_j^{\mu_j} 
\cdots {a_l^\dagger}^{\nu_l}a_l^{\mu_l} +h.c.
\label{pt1}\end{eqnarray}
More concisely, it is
\begin{eqnarray}
{\hat W}^{PT}=:f(a_1^\dagger,a_1,a_2^\dagger, a_2\cdots 
a_{j-1}^{\dagger},a_{j-1},a_j,a_j^\dagger\cdots a_l, a_l^{\dagger}):
\label{rept2}\end{eqnarray}
The notation $:\cdots:$ indicates putting all the creation operators at left and 
annihilation operators at right for each mode in the  functional $f$. \\
{\bf Non-Gaussian states.} 
The two mode squeezed state is a typical continuous variable  state produced in laboratories..
Practically speaking, we can rarely have a pure squeezed state due to  the noise.
We formulate the following impure squeezed state from a noisy source :
\begin{eqnarray}
\rho=\int^{r_1}_{r_0}p(r)S(r)|00\rangle \langle  00|S^\dagger(r)
{\rm d}r+\sum_{n_1,n_2=0} ^{M_1,M_2}p'(n_1,n_2)|n_1,n_2\rangle \langle  n_1,n_2|\label{wstate}.
\end{eqnarray}
Here $p(r)$ and $p'(n_1,n_2)$ are the probabilistic distribution functions satisfying
$\int p(r){\rm d}r+\sum_{n_1,n_2=0} ^{M_1,M_2}p'(n_1,n_2)=1,$ squeezing
operator $S(r)=\exp [r(a_1^\dagger a_2^\dagger-a_1a_2)]$ ($r > 0$), $r_0$ and $r_1$ are positive
numbers($r_1> r_0$). Using the normally ordered form of squeezing operator
we can simplify the integration part. Using Eq. (\ref{rept2}) we may perform the PT operation and we obtain 
\begin{eqnarray}
\rho^{PT}=\int p(r)\frac{1}{\cosh^2 r}\sum_{n_1,n_2}\tanh^{n_1+n_2} r {a_1^\dagger}^{n_1} 
{a_2^\dagger}^{n_2}
|00\rangle \langle  00|a_1^{n_2}a_2^{n_1}{\rm d}r+
\sum_{n_1,n_2=0} ^{M_1,M_2}p'(n_1,n_2)|n_1,n_2\rangle \langle  n_2,n_1|
.\end{eqnarray}
Consider the following linear superposed state in Fock space
\begin{eqnarray}
|\Psi\rangle =\frac{1}{\sqrt 2}(|M, M+2\rangle -|M+2,M\rangle ),
\end{eqnarray}
and $M > M_1, M > M_2$. We have
\begin{eqnarray}
\langle  \Psi|\rho^{PT}|\Psi\rangle 
=-\int^{r_1}_{r_0}\frac{p(r)}{\cosh^2 r}\tanh ^{2M+2}r {\rm d}r <  0.
\end{eqnarray}
Thus  {\it state defined in Eq. (\ref{wstate}) 
 is inseparable}.
As far as we have known,  Eq. (\ref{wstate}) is a $new$ formula
for the impure squeezed state.
 This state is somewhat similar to the Werner
state\cite{werner3} for the discrete system. However, the result is different. 
A mixture of Bell state can be 
separable very often, while a mixture of different squeezed vacuum and number state is always 
inseparable.

Besides the above noise due to the source, there are also  noises due to the channel.
We take  the phase damping channel as an example here. Suppose initially we have a perfect pure squeezed
state $S|00\rangle \langle  00|S^\dagger$. After solving the Master equation one obtains\cite{hiro} 
\begin{eqnarray}
\rho(t)=\frac{1}{\cosh^2r}\sum_{n_1,n_2}\tanh^{n_1+n_2}
r\exp(-\gamma t|n_1-n_2|^2)|n_1,n_1\rangle \langle  n_2,n_2|.
\end{eqnarray}
The entanglement formation of this $\rho(t)$ has been numerically calculated\cite{hiro}. 
However, as mentioned in \cite{hiro}, 
$"$it is not clear from the present numerical analysis whether the state is
always entangled for finite $\gamma t$$"$. The issue  can be easily
resolved in  Fock space. 
According to Ref. \cite{hor3}, if operator $\Omega={\rm tr}_D\rho\otimes I-\rho$ is not positive definite,
 there is always a scheme
to distill state $\rho$. 
We find here ${\rm tr}_D\rho(t)=\frac{1}{\cosh^2 r}\sum_n\tanh^{2n}|n\rangle \langle  n|$. It is easy 
to see $\langle  \Psi_2|\Omega|\Psi_2\rangle <  0$,
taking $|\Psi_2\rangle =\frac{1}{2}(|00\rangle +|11\rangle )$. Thus we draw the conclusion: 
{\it A two mode squeezed state is always distillable( and inseparable) in phase
damping channel.} \\
{\bf Gaussian states.}
 We define the Gaussian state in the following normally ordered form in Fock space.
\begin{eqnarray}\label{gauss}
\rho=:\exp\left[\frac{1}{2}\left(a R \widetilde a + a B a^\dagger
+\widetilde{a^\dagger} B^T \widetilde a+ \widetilde{a^\dagger} R^\dagger {a^\dagger}\right)\right]
:.
\end{eqnarray}
Here $a$ and $\widetilde {a^\dagger}$ are row vectors with $l$ components, e.g.
$a=(a_1,a_2\cdots a_l)$; $\widetilde {a^\dagger}=(a^\dagger_1,a^\dagger_2\cdots a_l^\dagger)$
 while $\widetilde a$ and $a^\dagger$ are $l$ component column vectors. Given a Gaussian operator
that is not in the normally ordered form(e.g., the squeezed thermal state\cite{wang0}), 
one is always  able to transform it into normally ordered form. 
There are a lot
of well developed methods to do so(see e.g. \cite{wang,wang1}).

According to Eq. (\ref{pt1}),  the transposition operation to the particles with subscript from $j$ to 
$l$ changes the state into
\begin{eqnarray}\label{gpt1}
\rho^{PT}=:\exp\left[\frac{1}{2}(a,\tilde {a^\dagger}) \left(\begin{array}{cc}
R'& B'\\B'^T & R'^\dagger  \end{array}\right)\left(\begin{array}{c}
\tilde a\\ a^\dagger \end{array}\right)
\right]:,
\end{eqnarray}
where
\begin{eqnarray}\label{gpt2}
\left(\begin{array}{cc}
R'& B'\\B'^T & R'^\dagger  \end{array}\right)=\sigma \left(\begin{array}{cc}
R& B\\B^T & R^\dagger  \end{array}\right)\sigma
\end{eqnarray}
and
$\sigma$ is a $2l\times 2l$ matrix defined as
$\sigma_{ff}=1$,
$
\sigma_{l+f,l+f }=1
$
for $f\leq j$;
$
\sigma_{f, l+f}=1
$,
$
\sigma_{l+f, f}=1
$
for $f>  j$; and all the other matrix elements are $0$. 
With the above  formulae, one can take PT operation to any Gaussian state very
easily.  To check the inseparability, we need check the positivity of the Gaussian
operator $\rho^{PT}$ 
and we have
\\{\bf Proposition }: {\it For the Gaussian operator $\rho^{PT}$ defined above, 
it is positive definite if
and only if all the eigenvalues of matrix $B'$ not less than $-1$}.\\
Proof: If we find certain operator $A$ so that $\rho^{PT}=AA^\dagger$, then $\rho^{PT}$ is
positive definite. 
Note that in the normally ordered form, operator $:\exp\left[(a B'^T a^\dagger
+\widetilde{a^\dagger} B' \widetilde a)/2\right]:=:\exp[\widetilde{a^\dagger}B'\tilde a]:$. 
Furthermore, if $B'> -1$, we can use the formula $$:\exp[(a B'^T a^\dagger
+\widetilde{a^\dagger} B' \widetilde a)/2]:=\exp[\widetilde{a^\dagger} \ln(B'+1) \widetilde a].$$ Denoting 
$A=\exp\left[\widetilde{a^\dagger} 
R'^\dagger {a^\dagger}/2\right]\exp\left[\frac{1}{2}\widetilde{a^\dagger} \ln(B'+1) \widetilde a\right] $ we have $\rho^{PT}=AA^\dagger$.
If some of the eigenvalues in matrix $B'$ are equal to $-1$, considering the fact that 
$:\exp(-a_i^\dagger a_i):=|0_i\rangle \langle  0_i|$( $|0_i\rangle $ is the vacuum state of the $i$th mode) and
$\sqrt{|0_i\rangle \langle  0_i|}=|0_i\rangle \langle  0_i|$, we can also easily factorize $\rho^{PT}$ into product
form  of certain $A$ and $A^\dagger$. Suppose the first $h$ eigenvalues 
of matrix $B'$ are $-1$, then we have 
\begin{eqnarray}
A=\exp\left[\widetilde{a^\dagger} 
R'^\dagger {a^\dagger}/2\right] U
:\exp [(\widetilde{a^\dagger} \Lambda \widetilde a)/2]:,
\label{gau3}\end{eqnarray}
where $\Lambda_{ii}=-1$ if $i\leq h$, $\Lambda_{ii}>  -1$ if $i> h$, and unitary operator
$U$ satisfies $ U
:\exp [a \Lambda a^\dagger
+\widetilde{a^\dagger} \Lambda' \widetilde a]:U^\dagger =$$:\exp[a B'^T a^\dagger
+\widetilde{a^\dagger} B' \widetilde a]:$. The unitary operator $U$ can be constructed in 
the following way:\\
Suppose matrix $B'$ can be diagonalized by unitary matrix $D$, i.e. $B'=D\Lambda D^{-1}$. We define $U$
to be the rotating operator satisfying $U \widetilde {a^\dagger} U^{\dagger}
=\widetilde {a^\dagger} D
=\widetilde {b^\dagger}$.
Then we have 
$$:\exp[\widetilde {a^\dagger} B' \widetilde a]:
=\sum_{k=0}^\infty\sum_i^n \frac{\lambda_i^k}{k!} {b_i^\dagger}^k b_i^k
=U:\exp(\widetilde{a^\dagger}\Lambda \widetilde a):U^\dagger.$$
Using the fact $\sqrt{:\exp(\widetilde{a^\dagger} \Lambda \widetilde{a}:)}=:\exp(\widetilde{a^\dagger} \Lambda \widetilde{a}/2):$  
we obtain Eq. (\ref{gau3}).

In the case that matrix $B'$ has any eigenvalue less than $-1$, 
without loss of any generality, we suppose $\Lambda_{11}<  -1$. Then the 
expectation value on state  $|\psi\rangle =\exp 
[-a R' \widetilde a/2] U|1,0,0\cdots 0\rangle $ is
\begin{eqnarray}
\langle  \psi|\rho^{PT}|\psi\rangle  
=\langle   1|:\exp[\Lambda_{11}a_1^\dagger a_1]: |1\rangle \cdot 
\langle  0|\exp[\Lambda_{22}a_2^\dagger a_2]|0\rangle \cdots \langle  0|\exp[\Lambda_{ll}a_l^\dagger a_l]|0\rangle .
\end{eqnarray}
Taking  Taylor expansion of the exponential we know the terms of expectation values
on vacuum state is $1$, while the expectation value on state $|1\rangle $ is
\begin{eqnarray}
\langle  1|:\exp[\Lambda_{11}a_1^\dagger a_1]: |1\rangle =\langle  1|(1+\Lambda_{11}a_1^\dagger a_1
+\frac{1}{2}\Lambda_{11}^2 {a_1^\dagger}^2a_1^2
-\cdots)|1\rangle =1+\Lambda_{11}<  0.
\end{eqnarray}
So operator $\rho^{PT}$ is not positive definite in this case. \\
With the proposition  one can easily check the 
inseparability of any Gaussian state. In general, one can easily 
obtain a sufficient condition for the inseparability of arbitrary state with them.
In particular, it has been shown\cite{wern} recently 
that for $N\otimes 1$ Gaussian state, 
there is no inseparable state that is positive definite after the PT operation. That is to say,
our proposition  have actually given the necessary and 
sufficient condition for the inseparability of bipartite Gaussian state which has $N$ 
particles in one subspace and $1$ particle in the other subspace.  Moreover, connected with 
theorem 1 in Ref. \cite{dist}, the proposition  gives the necessary and sufficient condition
for the distillability for all bipartite Gaussian states.
\\To demonstrate the calculation, we consider an un-normalized 3-variable Gaussian state defined as: 
\begin{eqnarray}
\rho =:\exp \left[\sum_{i<j}^3(\zeta_{ij} a_i^\dagger a_j^\dagger +{\zeta_{ij}}^*a_i a_j)  
-\lambda a_1^\dagger a_1-\lambda a_2^\dagger a_2-\lambda_3 a_3^\dagger a_3\right]:
\end{eqnarray}
and $0\leq\lambda\leq$ 1, $0\leq\lambda_3\leq$ 1. 
If $\rho$ is a physics state there are some restrictions to the  
parameters. Here we assume the parameters in $\rho$ satisfy all those restrictions.
Suppose mode $1$ and  mode $2$ are in subspace $A$ while mode $3$ are in subspace $B$. 
Taking PT operation to mode $3$ one easily obtains the matrix $B'$ as defined by equation(\ref{gpt2}).
The smallest eigenvalue($\omega$) of matrix $B'$ for this state can be caculated easily.
 The condition for $\omega <  -1$ is
\begin{eqnarray}
|\zeta_{13}|^2+|\zeta_{23}|^2 >  (\lambda+1)(\lambda_3+1). 
\end{eqnarray}
Our proposition here is presented in Fock space. 
It can be easily applied to a type of more general problem, the entanglement property in
a subspace of 
a global Gaussian state.
The application background for this question is broad.
For example, a multi particle bipartite state is prepared and then some of the particles are lost
due to the operational errors, or some of the particles there are not accessible due to any other
causes. We want to know whether the state in the accessible subspace  is still distillable.
For another example,  the initially prepared Gaussian state interacts
with its environment, a heat bath of harmonic oscillators. Finally we have
 a Gaussian state in the total system. But  only a subspace is accessible.
 It is interesting to judge whether the state in the accessible subspace is distillable. 
The whole task is easy with our proposition in Fock space, because it is easy to obtain
the explicit formula for a subspace state in a global Gaussian state, provided the global Gaussian 
state is 
expressed in the normally ordered form in Fock space.  

{\bf Bipartite state in a subspace of a global Gaussian state}
An $l'-$mode  global Gaussian state $\rho_{CD}$ can be defined in the form of equation (\ref{gauss}).  
Modes numbered from $1$ 
to $l$ belong to subspace $C$( which consists of subspaces $A$ and $B$),
  modes numbered from $l+1$ to $l' $ belong to the subspace
$D$. 
We have
\begin{eqnarray}
\rho_C={\rm tr_D}\rho_{CD}={\rm tr_D}
:\exp\frac{1}{2}\left[\beta_C M_C \widetilde{\beta_C}+
\beta_D M_D \widetilde{\beta_D}
+2\beta_D M_O\widetilde{\beta_C}
\right]:,
\end{eqnarray}
$\beta_C=(a_1,a_2\cdots a_{l},a_1^\dagger,a_2^\dagger\cdots a_l^\dagger)$, 
$\beta_D=(a_{l+1},a_{l+2}\cdots a_{l'},a_{l+1}^\dagger,a_{l+2}^\dagger\cdots a_{l'}^\dagger)$,
 $M_{C,D,O}=\left(\begin{array}{cc}
R_{C,D,O}& B_{C,D,O}\\B_{C,D,O}^T & R_{C,D,O}^\dagger  \end{array}\right)$  , 
matrix $R_C(B_C)$ is in $l\times l$ form, its matrix element
$R_{Cxx'}(B_{Cxx'})  =R_{xx'}(B_{xx'}) , 1\le x,x'\le l$,  
matrix $R_D(B_D)$ is in $(l'-l)\times (l'-l)$ form, its matrix element $R_{Dyy'}(B_{Dyy'})  
=R_{l+y,l+y'}(B_{l+y,l+y'}) , 1\le y,y'\le (l'-l)$,  
matrix $R_O$($B_O$) is in the
form $(l'-l)\times l$, its matrix element is 
$R_{Oz,z'}(B_{Oz,z'})  =R_{z,l+z'}(B_{z,l+z'}) , 1\le z\le l, 1\le z'\le l'-l$.
To take the partial trace in subspace $D$, we introduce the $l'-l$ mode coherent state $|Z_D>$
in subspace $D$, $|Z_D>=|z_{l+1},z_{l+2}\cdots z_{l'}>$, $a_{l+y}|Z_D>=z_{l+y}|Z_D>$. 
We have 
\begin{eqnarray}
\rho_C=:\int \exp\frac{1}{2}\left[\beta_C M_C \widetilde{\beta_C}+
Z_D^* M_D \widetilde{Z_D}
+2\beta_C M_O^T
\widetilde{Z_D}
\right]\prod_{t=j+1}^{l}{\rm d^2} z_t/\pi:
\end{eqnarray} 
Using the complex 
Gaussian integration
formula\cite{berzin} 
we obtain( up to a constant factor number): 
\begin{eqnarray}\label{pg}
\rho_C=: \exp\frac{1}{2}\left[\beta_C (M_C+2M_O^T  M_D^{-1}  M_O )
 \widetilde{\beta_C}
\right] :.
\end{eqnarray}
We can easily recast equation(\ref{pg}) into the form of equation(\ref{gauss}) through
$$
M_D^{-1}=\left(\begin{array}{cc}
(R_D-B_D{R_D^\dagger}^{-1}B_D^T)^{-1} & R_D^{-1}B_D(B_D^TR_D^{-1}B_D-{R_D^\dagger})^{-1}\\
{R_D^\dagger}^{-1}B_D^T(B_D{R_D^\dagger}^{-1}B_D^T-R_D)^{-1} & 
(R_D^\dagger-B_D^TR_D^{-1}B_D)^{-1}
\end{array}\right).
$$
So we have given the explicit Gaussian formula for the state in a subspace of  a global 
Gaussian state. This equation, together with the known criterions on distillability and 
inseparability of Gaussian states gives the correspondings criterions  in
 a subspace of a global
Gaussian state.
  
{\bf Concluding remark.} In summary, we have presented a general method to detect the
inseparability and distillability for states in Fock space.   
We have shown the inseparability for some non-Gaussian states.
We give the necessary and sufficient condition for positivity of 
the normally ordered multi-mode Gaussian operator. This is also 
the necessary and sufficient
condition for the distillability of all bipartite Gaussian states and 
 the necessary and sufficient condition for the inseparability
of  $N\otimes 1$ Gaussian states. With the explicit formula for the bipartite state 
in a subspace of a Gaussian state given by us, all these criterions are applicable
to a subspace inside the Gaussian state. 

{\bf Acknowledgement}: We thank Prof. Imai Hiroshi for support. We thank Dr. M. Hachimori   and
Dr. W. Y. Hwang for helpful discussions.

\end{document}